\documentclass[%
 reprint,
 superscriptaddress,
 groupedaddress,
 showpacs,preprintnumbers,
 nobibnotes,
 amsmath,amssymb,
 aps,
 floatfix,
]{revtex4-1}

\usepackage{graphicx}

\usepackage{color}
\usepackage{hyperref}

\begin{document}

\title{
Doping-induced spin-orbit splitting in Bi-doped ZnO nanowires
}

\author{Mehmet Aras}
\author{S\"{u}meyra G\"{u}ler-K{\i}l{\i}\c{c}}\email{sumeyra@gtu.edu.tr}
\author{\c{C}etin K{\i}l{\i}\c{c}}\email{cetin\_kilic@gtu.edu.tr}
\affiliation{Department of Physics, Gebze Technical University, 41400 Gebze Kocaeli, Turkey}


\begin{abstract}
\centerline{\sl Published version available at \url{https://doi.org/10.1103/PhysRevB.95.155404}}
\vspace*{6pt}
Our predictions, based on density-functional calculations,
  reveal that
  surface doping of ZnO nanowires with Bi leads to a linear-in-$k$ splitting of the conduction-band states,
  through spin-orbit interaction,
  due to the lowering of the symmetry in the presence of the dopant.
This finding implies that spin-polarization of the conduction electrons in Bi-doped ZnO nanowires
  could be controlled with applied electric (as opposed to magnetic) fields,
  making them candidate materials for spin-orbitronic applications.
Our findings also show that the degree of spin splitting
  could be tuned by adjusting the dopant concentration.
Defect calculations and \textit{ab initio} molecular dynamics simulations
  indicate that \textit{stable} doping configurations exhibiting the foregoing linear-in-$k$ splitting
  could be \textit{realized} under reasonable thermodynamic conditions.
\end{abstract}


\maketitle

\section{Introduction}

The use of quantum wires in spintronic applications
  is enabled by engineering their inversion asymmetries
  in the presence of strong spin-orbit (SO) interactions\cite{moroz99,zhang06,pramanik07,quay10,luo11,crepaldi13,park13},
  which lifts the Kramers degeneracy of the electronic states.
Although \textit{doping} could clearly be employed as a means
  for breaking the inversion symmetry,
  it remains largely unexplored
  \textit{if} the spintronic functionalities of a quasi-one-dimensional material
  could be extended by doping with a heavy element whose presence usually enhances the SO coupling.
The present study is devoted to exploring the latter,
  where doping non-magnetic ZnO nanowires with the heavy element Bi
  is taken as an exemplar.
Our interest in ZnO nanowires stems from
  the fact that doping them with a variety of elements is practicable,
  leading to emergent new functionalities.
Controlled doping of ZnO nanowires has indeed led to diverse applications in recent years such as
  photodetectors\cite{hsu14},
  p-n homojunction rectifiers\cite{li13},
  sensors\cite{spencer12},
  light-emitting diodes\cite{lupan11},
  piezoelectric generators\cite{lu09},
  field-effect transistors\cite{yuan08}, and
  field emitters\cite{huang07}.
Likewise, doping ZnO nanowires with transition metals\cite{chang03,cui05,ruiz11}
                                 or rare-earth elements\cite{iqbal09,ma12}
  makes them candidate building blocks in bottom-up assembly of spintronic devices.
To add to the foregoing cases, 
  we will show that the functional properties of ZnO nanowires
  could further be diversified by doping with Bi,
  making them candidate materials for spin-orbitronic\cite{hoffmann15} applications.
It should be remarked that
  Bi-doped ZnO nanowires have not been explored prior to this research as regards the spintronic properties,
  whose electrical properties have nevertheless been characterized
  in experimental\cite{xu07} and computational\cite{kilic16}
  studies
  which agreed that
  bismuth acts as a \textit{donor} in ZnO nanowires,
  despite its \textit{acceptor} behavior in ZnO thin films\cite{xiu06} and varistors\cite{smith89}.

We find,
  via noncollinear density-functional calculations that take into account the SO interaction, that
  the presence of bismuth
  as a \textit{substitutional} dopant in a semiconducting ZnO nanowire
  leads to a linear-in-$k$ spin-orbit splitting of the conduction band (CB) states.
Since the latter
  facilitates phenomena involving electric-field- and current-induced 
  spin polarization of semiconductor electrons\cite{silsbee04,hoffmann15},
  our finding implies that
  Bi-doped ZnO nanowires could be utilized in spintronic applications.
Our analysis reveals that
  the foregoing SO splitting
  arises from the interaction of the CB electrons (donated by the dopant)
  with the effective spin-orbit field $\mathbf B_{\text{\tiny SO}}$
  which originates mostly from the inhomogeneous electric potential gradient of the host nanowire.
It is interesting to note that
  this splitting
  could \textit{not} be attributed to the Dresselhaus\cite{dresselhaus55} and 
                                               Rashba\cite{bychkov84} interactions,
  the latter being customarily invoked to describe
  the $k$-dependent spin splitting
  in quasi-one-dimensional systems\cite{moroz99,zhang06,pramanik07,quay10,luo11,crepaldi13,park13}.
Thus,
  neither bulk- nor structural-inversion asymmetries\cite{dresselhaus55,bychkov84}
  of the host nanowire alone give rise to the SO-induced effects explored here,
  which indeed \textit{cease} to exist in the \textit{absence} of the dopant.

We discuss our findings in detail in Sec.~\ref{sonuclar},
  following a description of our computational modeling and simulation framework in Sec.~\ref{yontem},
  and present a brief summary in Sec.~\ref{netice}.
In the Appendix, we explore the lowering of the symmetry in the presence of the dopant.

\section{\label{yontem}Computational modeling and simulations}

We performed geometry optimizations,
               electronic structure calculations,
               defect calculations, and
               finite-temperature \textit{ab initio} molecular dynamics (MD) simulations
  for a variety of periodic supercells
  that contain a host (ZnO) nanowire and one or two Bi atoms as dopant and/or adatom,
  using the Vienna \textit{ab initio} simulation package\cite{kresse96} (VASP) and
  adopting the rotationally-invariant DFT+$U$ approach\cite{dudarev98}
  in combination with the Perdew-Burke-Ernzerhof exchange-correlation functional\cite{perdew96}.
We employed the projector augmented-wave method\cite{blochl94,kresse99},
  treating the 2$s$ and 2$p$,
               3$d$ and 4$s$, and
               6$s$ and 6$p$
  states as valence states for oxygen,
                               zinc, and
                               bismuth, respectively.
Plane wave basis sets were used to represent the electronic states, which were determined
    by imposing a kinetic energy cutoff of 400 eV.
Increasing the latter by 10~\% resulted in a variation smaller than 0.5~\%
  in the SO splitting energies reported in Sec.~\ref{sonuclar}.
The Hubbard $U$ was applied only to the Zn 3$d$ states,
  whose value was set to 7.7 eV\cite{kilic16}.
The DFT+$U$ approach was preferred over the standard (semilocal) DFT calculations
  because the inclusion of $U$ improves the underestimation of the $d$ state binding energies\cite{aras14},
  and therefore reduces the spurious $pd$ hybridization in the upper valence band of ZnO.
In geometry optimizations,
  ionic relaxations were performed for each atomic structure
  to minimize the total energy $E$,
  until the maximum value of residual forces on atoms was reduced to be smaller than $10^{-2}$~eV/\AA,
  using the $\Gamma$-point for sampling the supercell Brillouin zone (BZ).
Note that the BZ sampling is virtually achieved through \textit{zone folding}
  since we in practice use supercells made of $n$ unit cells, as described in the next paragraph, with $n>4$.
To obtain an error bar for $E$ owing to the BZ sampling,
  we performed a number of test calculations using the primitive unit cell of a host nanowire and
  increasing the number of $\mathbf k$-points from 4 to 11,
  which showed a variation in the energy (per ZnO unit) smaller than 0.2 meV.
In electronic structure calculations, spin-orbit coupling was taken into account
   by utilizing the noncollinear mode of VASP\cite{hobbs00,marsman02}.
Convergence criterion for the electronic self-consistency was set up to
  10$^{-6}$~eV (10$^{-8}$~eV) in geometry optimizations (electronic structure calculations).
\textit{Ab initio} MD simulations were performed at temperature 600~K
  with the aid of a Nos\'e-Hoover thermostat\cite{nose84},
  integrating the equations of motion via the Verlet algorithm with a time step of 1 fs.

We considered a variety of atomic configurations within the supercell approach,
  including the ZnO nanowire with
    (i) the Bi dopant substituting Zn at a surface site,
        denoted as $\text{[(ZnO)}_N\text{]}_n\text{:Bi}_{\text{Zn}}$,
   (ii) the Bi adatom,
        denoted as $\text{[(ZnO)}_N\text{]}_n\text{+Bi}$, and
  (iii) the substitutional Bi dopant together with a Bi adatom,
        denoted as $\text{[(ZnO)}_N\text{]}_n\text{+Bi}\text{:Bi}_{\text{Zn}}$.
The primitive unit cell for a host nanowire,
  which contains $\text{(ZnO)}_N$,
  is cut from bulk ZnO in wurtzite structure
  in such a way that the wire axis coincide with the [0001] direction of wurtzite,
  and therefore has a hexagonal cross section.
A host nanowire $\text{[}\text{(ZnO)}_N\text{]}_n$ is constructed
  by combining $n$ consecutive unit cells;
  the supercell length $L$ along the wire axis
  is given by $L$=$nc$
  where $c$ denotes the length of periodicity along the wire axis.
The symmetries of the host nanowires employed in this study
  are described by the R 70 (p6$_3$mc) rod group\cite{tronc06},
  which means that the undoped $\text{[}\text{(ZnO)}_N\text{]}_n$ nanowire
  has the same (C$_{6\text{v}}$) point group as bulk (wurtzite) ZnO.
The incorporation of the Bi dopant and/or adatom
  eliminates these symmetries since
  \textit{no} symmetry elements are present in 
  $\text{[(ZnO)}_N\text{]}_n\text{:Bi}_{\text{Zn}}$,
  $\text{[(ZnO)}_N\text{]}_n\text{+Bi}$, and
  $\text{[(ZnO)}_N\text{]}_n\text{+Bi}\text{:Bi}_{\text{Zn}}$.

As for the adatom configurations,
  we computed
  the adsorption energy $E_{\text{ad}}=E(\text{nanowire}\text{+Bi})-E(\text{nanowire})-E(\text{Bi})$,
  where $E(\text{Bi})$ denotes the energy of the Bi atom,
  and determined the \textit{lowest-energy} adatom configuration.
As for the doping configurations,
  we computed the formation energy
  as a function the bismuth chemical potential $\mu_{\text{\tiny Bi}}$
  under Zn--\textit{rich} and Zn--\textit{poor} conditions,
  assuming a chemical equilibrium between the doped nanowire and the reservoirs of its constituent atoms.
The formation energy is given by
  $\Delta H_{\text{f}}=[E(\text{nanowire}+m\text{ Bi})-E(\text{nanowire})+E(\text{Zn})-mE(\text{Bi})]+\mu_{\text{\tiny Zn}}-m\mu_{\text{\tiny Bi}}$,
  where $m$ denotes the number of Bi atoms transferred from the bismuth reservoir to the doped nanowire,
   \textit{i.e.}, $m$=1 for $\text{\rm [(ZnO)}_{24}\text{]}_5\text{:Bi}_{\text{Zn}}$ and
         $m$=2 for $\text{\rm [(ZnO)}_{24}\text{]}_5\text{+Bi}\text{:Bi}_{\text{Zn}}$.
Here $E(\text{Zn})$ denotes the energy per atom of bulk Zn metal,
  which is set as the zero of the zinc chemical potential $\mu_{\text{\tiny Zn}}$.
Zn--\textit{rich} and Zn--\textit{poor} conditions
  correspond respectively to $\mu_{\text{\tiny Zn}}=0$ and $\mu_{\text{\tiny Zn}}=\Delta H_N$,
  where $\Delta H_N$ denotes the \textit{heat of formation} (per formula unit) of the nanowire 
  made of $N$ Zn--O pairs,
  due to the chemical equilibrium between the nanowire and
  the reservoirs of its constituent atoms.
Note that the latter sets the range of values for the oxygen chemical potential $\mu_{\text{\tiny O}}$ as well,
  since $\mu_{\text{\tiny Zn}}+\mu_{\text{\tiny O}}=\Delta H_N$.

\section{\label{sonuclar}Results and Discussion}

\begin{figure*}
\centering
  \includegraphics[width=0.723\textwidth]{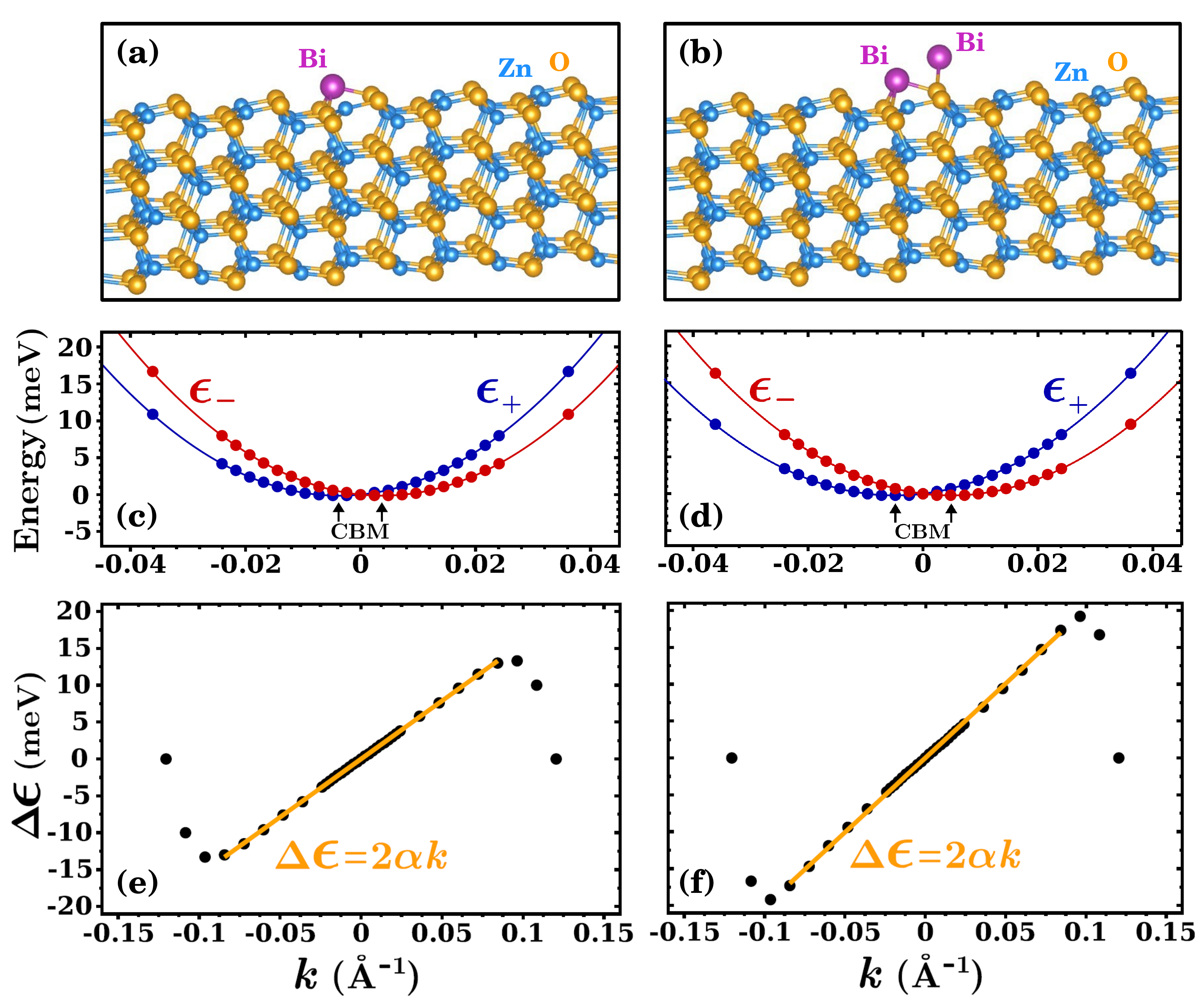}
  \caption{Optimized atomic structures for (a) $\text{[(ZnO)}_{24}\text{]}_5\text{:Bi}_{\text{Zn}}$ and
                                           (b) $\text{[(ZnO)}_{24}\text{]}_5\text{+Bi}\text{:Bi}_{\text{Zn}}$.
          [(c) and (d)] SO-split conduction bands for structures in (a) and (b), respectively.
          [(e) and (f)] Plots showing the variation of the splitting energy $\Delta \epsilon$ with the wave vector $k$.
          }
  \label{f:cb}
\end{figure*}

\begin{table}[b]
\caption{\label{t:fitband}
         The values for the effective mass $m^{\ast}$ (in free-electron mass),
                        the coefficient $\alpha$ (in eV\AA), and
                        the expectation value $\langle \mathbf m \rangle_\pm$ of the magnetization density (in Bohr magneton).
        }
\begin{ruledtabular}
\begin{tabular}{lccc}
System                                                               & $m^{\ast}$ & $\alpha$   & $\langle \mathbf m \rangle_\pm$         \\ \hline
  $\text{[(ZnO)}_{24}\text{]}_5\text{:Bi}_{\text{Zn}}$           & 0.36       & 0.080      & ($\mp 0.11 $, $\pm 0.46 $, $\mp 0.03 $) \\
  $\text{[(ZnO)}_{24}\text{]}_5\text{+Bi}\text{:Bi}_{\text{Zn}}$ & 0.39       & 0.095      & ($\pm 0.04 $, $\pm 0.32 $, $\pm 0.34 $)
\end{tabular}
\end{ruledtabular}
\end{table}

For the optimized atomic structures
  displayed in Figs.~\ref{f:cb}(a) and \ref{f:cb}(b),
  viz. $\text{[(ZnO)}_{24}\text{]}_5\text{:Bi}_{\text{Zn}}$ and
       $\text{[(ZnO)}_{24}\text{]}_5\text{+Bi}\text{:Bi}_{\text{Zn}}$,
  the SO-split conduction bands
  are shown in Figs.~\ref{f:cb}(c) and \ref{f:cb}(d), respectively.
It is seen
  that the dispersion of the these bands 
  is \textit{accurately} described by the solid curves (in blue and red)
  that represent fits in the form
\begin{equation}
  \epsilon_{\pm}(k)=\frac{\hbar^2}{2m^{\ast}}k^2 \pm \alpha k,
  \label{e:disp}
\end{equation}
which resembles the Bychkov-Rashba expression\cite{bychkov84}.
The values for the effective mass $m^{\ast}$ and 
               the linear coefficient $\alpha$ are given in Table~\ref{t:fitband}
  where the expectation values $\langle \mathbf m \rangle_\pm$
    of the magnetization density $\mathbf m(\mathbf r)$
    corresponding to the $\epsilon_\pm$ bands are also given.
Although these $\alpha$ values appear to be small                                        
  compared to the values of the Rashba parameter $\alpha_{\text{R}}$                     
  in ``giant'' Rashba systems such as Pt--Si nanowires\cite{park13},                     
  it should be remarked that                                                             
  the SO splitting here is obtained by adding a \textit{single} heavy atom               
  to a \textit{light} oxide,                                                             
  corresponding to a low ({\textit{i.e.}, one Bi atom per 120 Zn--O pairs) concentration.
This is appealing                                                                        
  because enhancing spin-orbit coupling is usually accomplished                          
  by using significant amounts of heavy elements.                                      
The $\alpha$ values in Table~\ref{t:fitband} are indeed of the same magnitude as         
  the $\alpha_{\text{R}}$ values in III-V heterostructures\cite{nitta97},                
  making us expect Bi-doped ZnO nanowires find use in spintronic devices.                
The linear $k$ dependence of the band energies $\epsilon_{\pm}$
  in the vicinity of the conduction band minimum (CBM)
  is further confirmed by plotting
  the splitting energy $\Delta \epsilon = \epsilon_{+}-\epsilon_{-}$ with respect to
  the wave vector $k$ in Figs.~\ref{f:cb}(e) and \ref{f:cb}(f).
The contribution from Bi to the states of the bands in Fig.~\ref{f:cb}(c) [Fig.~\ref{f:cb}(d)]
  is in the range of 3--4~\% [6--7~\%],
  cf. Fig.~S1 (see Ref.~\onlinecite{supmat} for Supplemental Material),
  showing that these bands are predominantly derived 
  from the respective conduction bands of the host (ZnO) nanowire.
Hence the states of the SO-split bands have more of character of \textit{extended} states
  that facilitate (spin-polarized) electrical conduction.

\begin{figure}
\centering
  \includegraphics[width=0.482\textwidth]{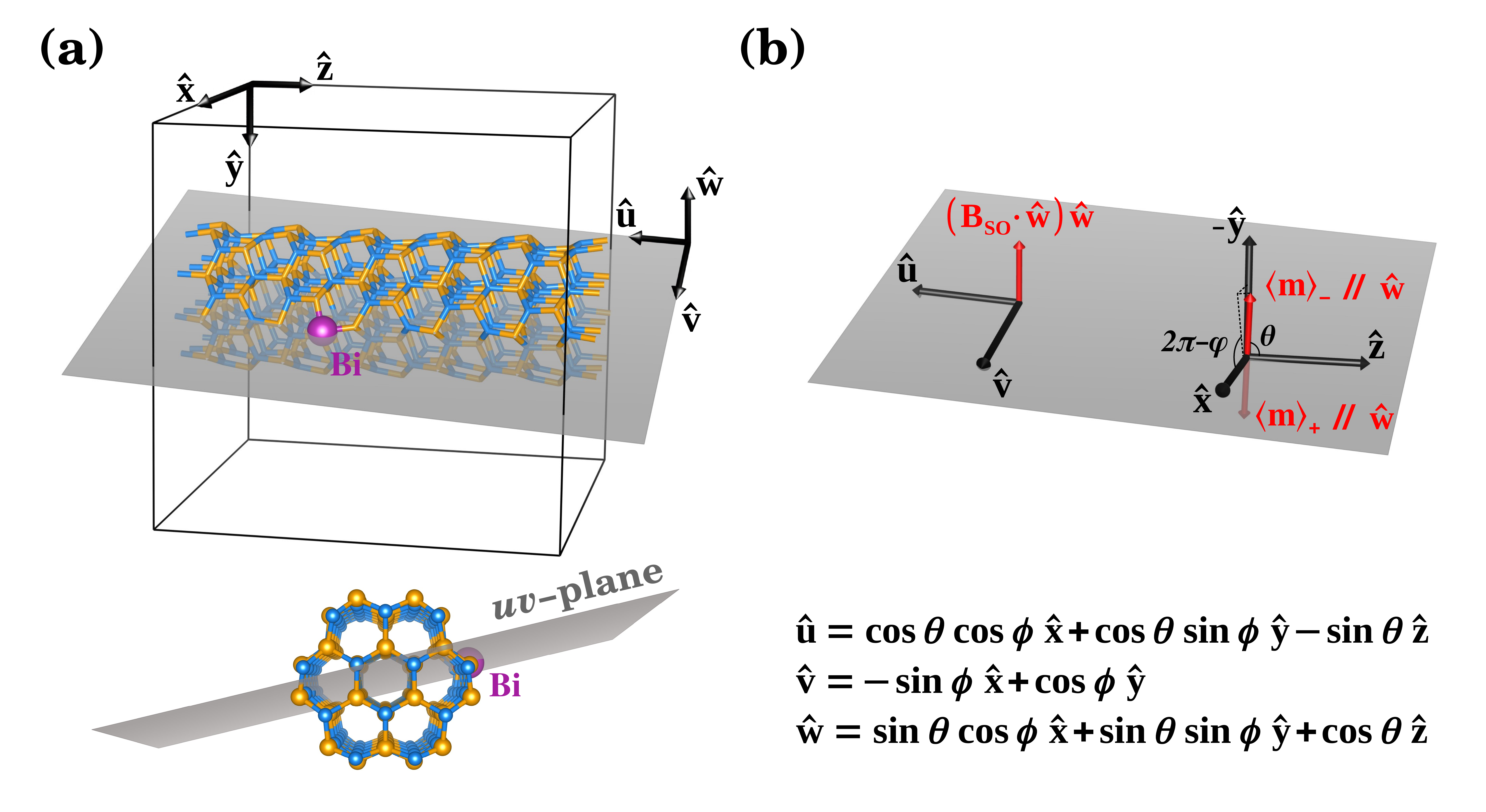}
  \caption{(a) The $uv$ plane that is perpendicular to $\langle \mathbf m \rangle_\pm$ and the $uvw$ coordinate system are shown
               for $\text{[(ZnO)}_{24}\text{]}_5\text{:Bi}_{\text{Zn}}$.
           (b) A drawing showing the $w$-component of ${\mathbf B}_{\text{\tiny SO}}$ 
               in the direction parallel to $\langle \mathbf m \rangle_-$.
          }
  \label{f:plane}
\end{figure}

\begin{figure*}
\centering
  \includegraphics[width=1.000\textwidth]{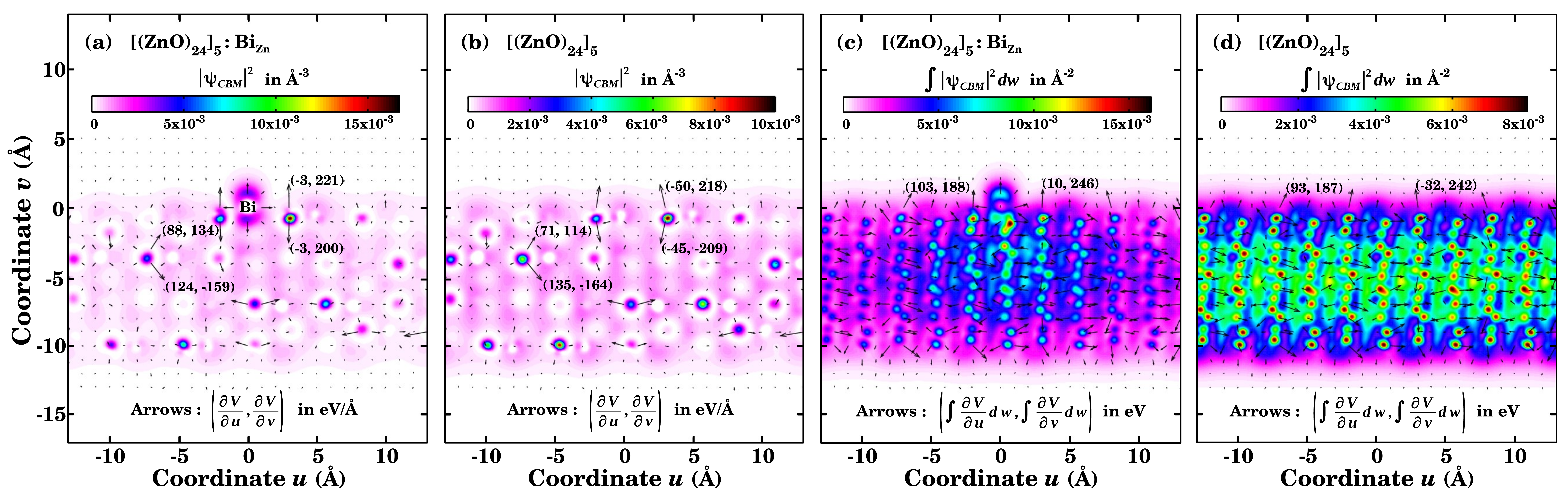}
  \caption{The arrows represent [(a) and (b)] the potential gradient on the $uv$ plane that is perpendicular to $\langle \mathbf m \rangle_\pm$
             and [(c) and (d)] the vector quantity ($\int \frac{\partial V}{\partial u} dw$, $\int \frac{\partial V}{\partial v} dw$)
           for [(a) and (c)] Bi-doped and [(b) and (d)] undoped ZnO nanowires.
           Colored contour plots of [(a) and (b)] the CBM state charge density $|\psi_{\text{\tiny CBM}}|^2$ and
                                    [(c) and (d)] the integrated quantity $\int |\psi_{\text{\tiny CBM}}|^2 dw$
             are superimposed.
          }
  \label{f:quiver}
\end{figure*}

Consider,
  for the purpose of interpretation,
  a Kramers doublet with two-component spinor wave functions
  $\psi_{\mathbf k}^-(\mathbf r) =
  \psi_{\mathbf k}(\mathbf r)
  \left [ \begin{array}{c} \cos{\frac{\theta}{2}} e^{-i\frac{\phi}{2}} \\
                           \sin{\frac{\theta}{2}} e^{ i\frac{\phi}{2}}
          \end{array} \right ]$
  and
  $ \psi_{\mathbf k}^+(\mathbf r) =
  \psi^\ast_{\mathbf k}(\mathbf r)
  \left [ \begin{array}{c} -\sin{\frac{\theta}{2}} e^{-i\frac{\phi}{2}} \\
                            \cos{\frac{\theta}{2}} e^{ i\frac{\phi}{2}}
          \end{array} \right ]$
where $\psi_{\mathbf k}(\mathbf r)=e^{i\mathbf k \cdot \mathbf r} u_{\mathbf k}(\mathbf r)$
  denotes the wave function of the respective Bloch state
  in the absence of the SO interaction.
The expectation value of the SO interaction operator\cite{davydov65}
  with $\psi_{\mathbf k}^-(\mathbf r)$
  can be expressed as
\begin{equation}
  \langle \psi_{\mathbf k}^- | H_{\text{\tiny SO}} | \psi_{\mathbf k}^- \rangle
  = -\langle {\mathbf m} \rangle_- \cdot 
     \langle \mathbf B_{\text{\tiny SO}}(\mathbf k) \rangle,
  \label{e:hso}
\end{equation}
where
$\langle \mathbf B_{\text{\tiny SO}}(\mathbf k) \rangle = \int \psi^\ast_{\mathbf k}(\mathbf r) \mathbf B_{\text{\tiny SO}} \psi_{\mathbf k}(\mathbf r) d^3r $
is the $\mathbf k$-dependent matrix element of the operator
$\mathbf B_{\text{\tiny SO}} = (-\vec{\nabla} V \times \mathbf p)/2emc^2$ (in the nonrelativistic limit).
A similar equation follows for 
  $\langle \psi_{\mathbf k}^+ | H_{\text{\tiny SO}} | \psi_{\mathbf k}^+ \rangle$,
  but for our purpose it suffices to focus on
  $\langle \psi_{\mathbf k}^- | H_{\text{\tiny SO}} | \psi_{\mathbf k}^- \rangle$.
According to Eq.~(\ref{e:hso}),
  a linear-in-$k$ SO splitting as in Eq.~(\ref{e:disp})
  occurs \textit{only if}
  (i)~$\langle {\mathbf m} \rangle_-$ is nonzero
  and
  (ii)~$\langle \mathbf B_{\text{\tiny SO}}(\mathbf k) \rangle$ has nonzero components
  in the direction parallel to $\langle {\mathbf m} \rangle_-$.
The condition (i) is satisfied
  for $\text{[(ZnO)}_{24}\text{]}_5\text{:Bi}_{\text{Zn}}$
  and $\text{[(ZnO)}_{24}\text{]}_5\text{+Bi}\text{:Bi}_{\text{Zn}}$
  as indicated by the $\langle {\mathbf m} \rangle_\pm$ values given in Table~\ref{t:fitband}.
In other words,
  the Bi incorporation leads to \textit{magnetization} of the two lowest CB states
  although the total magnetic moment $\mathbf M=\int \mathbf m(\mathbf r) d^3r $ is \textit{zero}
  for both $\text{[(ZnO)}_{24}\text{]}_5\text{:Bi}_{\text{Zn}}$ and 
           $\text{[(ZnO)}_{24}\text{]}_5\text{+Bi}\text{:Bi}_{\text{Zn}}$.
It remains to see if the condition (ii) is also satisfied.
It is convenient
  to use the $uvw$ coordinate system depicted in Figs.~\ref{f:plane}(a) and \ref{f:plane}(b)
  where $\hat{\mathbf w}$ denotes the unit vector in the direction of $\langle {\mathbf m} \rangle_-$.
Because of the dot product in Eq.~(\ref{e:hso}),
  we are interested only in the $w$-component of $\langle \mathbf B_{\text{\tiny SO}}(\mathbf k) \rangle$
  that is perpendicular to the $uv$-plane, cf. Fig.~\ref{f:plane}(b).
The latter,
  \textit{i.e.}, $\langle \mathbf B_{\text{\tiny SO}}(\mathbf k) \rangle \cdot \hat{\mathbf w}$,
  involves $\partial V/\partial u$ and $\partial V/\partial v$,
  and does \textit{not} have any term with $\partial V/\partial w$
  owing to the cross product in the definition of $\mathbf B_{\text{\tiny SO}}$.
In this line of reasoning,
  the origin of the SO splitting of the bands of $\text{[(ZnO)}_{24}\text{]}_5\text{:Bi}_{\text{Zn}}$,
  cf. Fig.~\ref{f:cb}(c),
  is sought  by plotting the potential gradient $\vec{\nabla} V$ on the $uv$ plane in Fig.~\ref{f:quiver}(a)
  where a colored contour plot of the state charge density $|\psi_{\text{\tiny CBM}}|^2$ is superimposed.
Respective plots for the undoped nanowire $\text{[(ZnO)}_{24}\text{]}_5$
  are given in Fig.~\ref{f:quiver}(b) for comparison.
The arrows in Figs.~\ref{f:quiver}(a) and \ref{f:quiver}(b)
  show the spatial inhomogeneity and asymmetry
  in the $\partial V/\partial u$ and $\partial V/\partial v$ components of the potential gradient,
  which gives rise to the effective field $\mathbf B_{\text{\tiny SO}}$, and thus
  implies the satisfaction of the aforementioned condition (ii).
The \textit{centrosymmetric} pattern of arrows around the Bi atom in Fig.~\ref{f:quiver}(a) should be noticed,
  indicating that the Bi atom would \textit{not} make an appreciable contribution to $\mathbf B_{\text{\tiny SO}}$.
The inhomogeneity of the potential gradient
  is further characterized
  in Figs.~\ref{f:quiver}(c) and \ref{f:quiver}(d)
  with the aid of integrated quantities
  ($\int \frac{\partial V}{\partial u} dw$, $\int \frac{\partial V}{\partial v} dw$) and
  $\int |\psi_{\text{\tiny CBM}}|^2 dw$.
A comparison of the arrows in Fig.~\ref{f:quiver}(a) [Fig.~\ref{f:quiver}(c)]
  with those in Fig.~\ref{f:quiver}(b) [Fig.~\ref{f:quiver}(d)]
  also indicates that the foregoing spatial inhomogeneity is \textit{not} induced by the dopant.
However, as will be elaborated in the Appendix,
  the incorporation of the dopant reduces the symmetry of the potential,
  rendering the CBM wave function to possess a lower symmetry.
If the symmetry of the latter was conserved (\textit{i.e.}, remained the same as in the undoped ZnO nanowire),
  the matrix element $\langle \mathbf B_{\text{\tiny SO}}(\mathbf k) \rangle$ in Eq.~(\ref{e:hso})
  would vanish, and no SO splitting could occur
  (which is clearly not the case).
Nonetheless,
  the change in the potential gradient due to the dopant is small,
  as quantified in the Appendix.
It can accordingly be said that
  the effective field $\mathbf B_{\text{\tiny SO}}$ originates
  mostly
  from the nonvanishing potential gradient of the host (ZnO) nanowire,
  owing to its noncentrosymmetric (wurtzite) structure.
In sum,
  on one hand the donation of electrons of the Bi dopant leads to magnetization of the CB states, and
  on the other hand,
  the inhomogeneity of the potential gradient of the host nanowire
  gives rise to an effective SO field $\mathbf B_{\text{\tiny SO}}$.
Considering a Zeeman-type interaction, according to Eq.~(\ref{e:hso}),
  between $\mathbf B_{\text{\tiny SO}}$ and the conduction electrons in the foregoing CB states
  explains the SO splitting of the bands shown in Figs.~\ref{f:cb}(c) and \ref{f:cb}(d).

\begin{figure*}
\centering
  \includegraphics[width=0.723\textwidth]{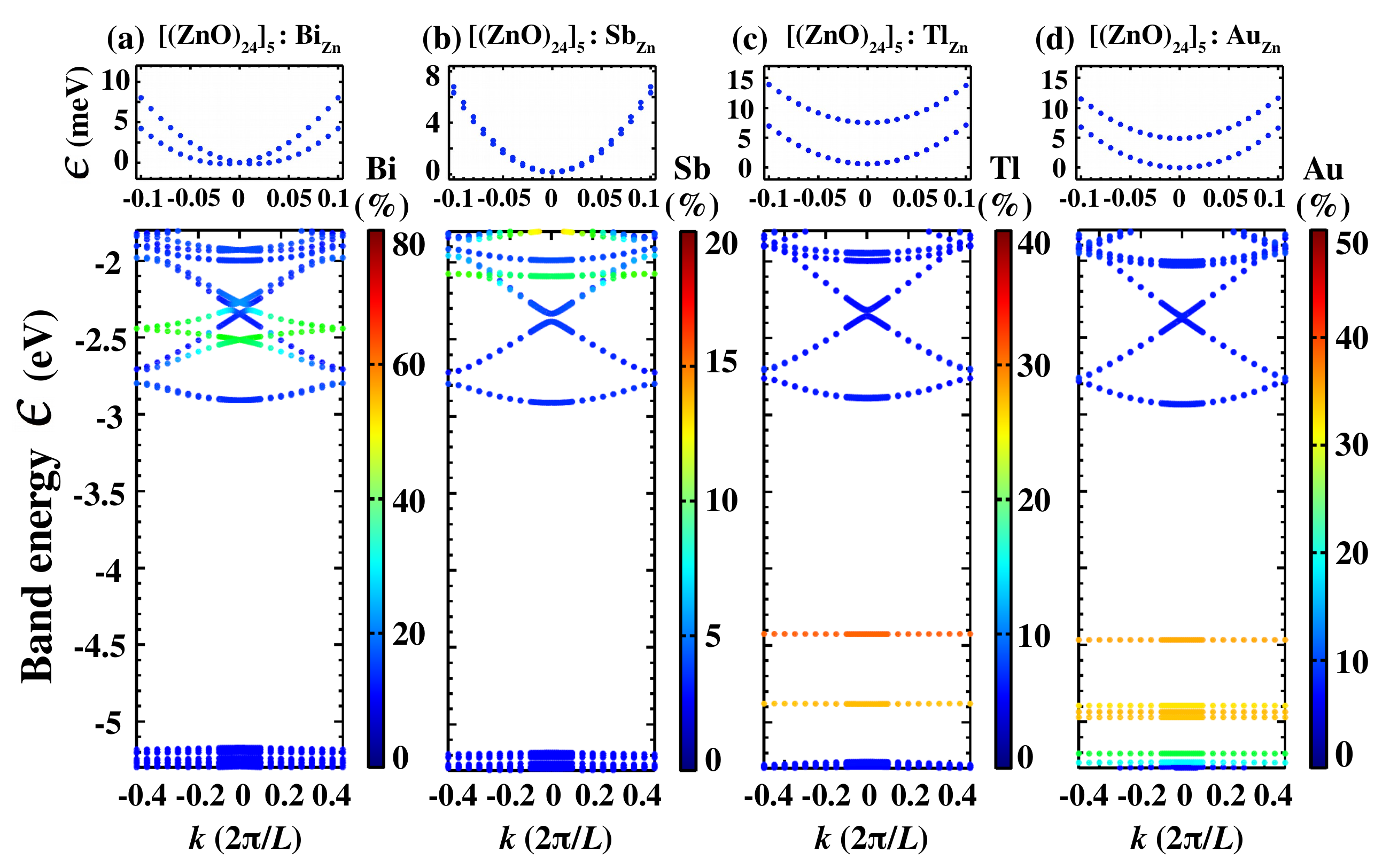}
  \caption{The electronic energy bands calculated for
           (a) $\text{[(ZnO)}_{24}\text{]}_5\text{:Bi}_{\text{Zn}}$,
           (b) $\text{[(ZnO)}_{24}\text{]}_5\text{:Sb}_{\text{Zn}}$,
           (c) $\text{[(ZnO)}_{24}\text{]}_5\text{:Tl}_{\text{Zn}}$, and
           (d) $\text{[(ZnO)}_{24}\text{]}_5\text{:Au}_{\text{Zn}}$.
           The circles are colored to reflect 
           the percent contribution from the dopant (Bi, Sb, Tl, or Au) to the electronic states.
           The upper panels show a close-up view of the \textit{two} lowest conduction bands.
          }
  \label{f:s2}
\end{figure*}

It is discernible in Figs.~\ref{f:quiver}(a) and \ref{f:quiver}(c) that
  the CBM wave function is slightly localized in the regions close to the dopant,
  which has nevertheless comparable contributions from the dopant and host atoms.
The latter precludes formation of
  a two-dimensional (2D) electron gas
  [although the lowest conduction bands in Fig.~\ref{f:cb}(c) are partially occupied],
  which means that
  the SO splitting in Bi-doped ZnO nanowire 
  does \textit{not} originate from the Rashba interaction\cite{bychkov84}.
It should also be pointed out that
  the band splitting in Figs.~\ref{f:cb}(c) and \ref{f:cb}(d)
  is \textit{not} caused by the Dresselhaus effect\cite{dresselhaus55}
  since the Dresselhaus spin splitting does \textit{not} occur 
  along the $k_z$ direction in wurtzite semiconductors\cite{de10}
  such as ZnO\cite{schleife09}.

\begin{figure}
\centering
  \includegraphics[width=0.482\textwidth]{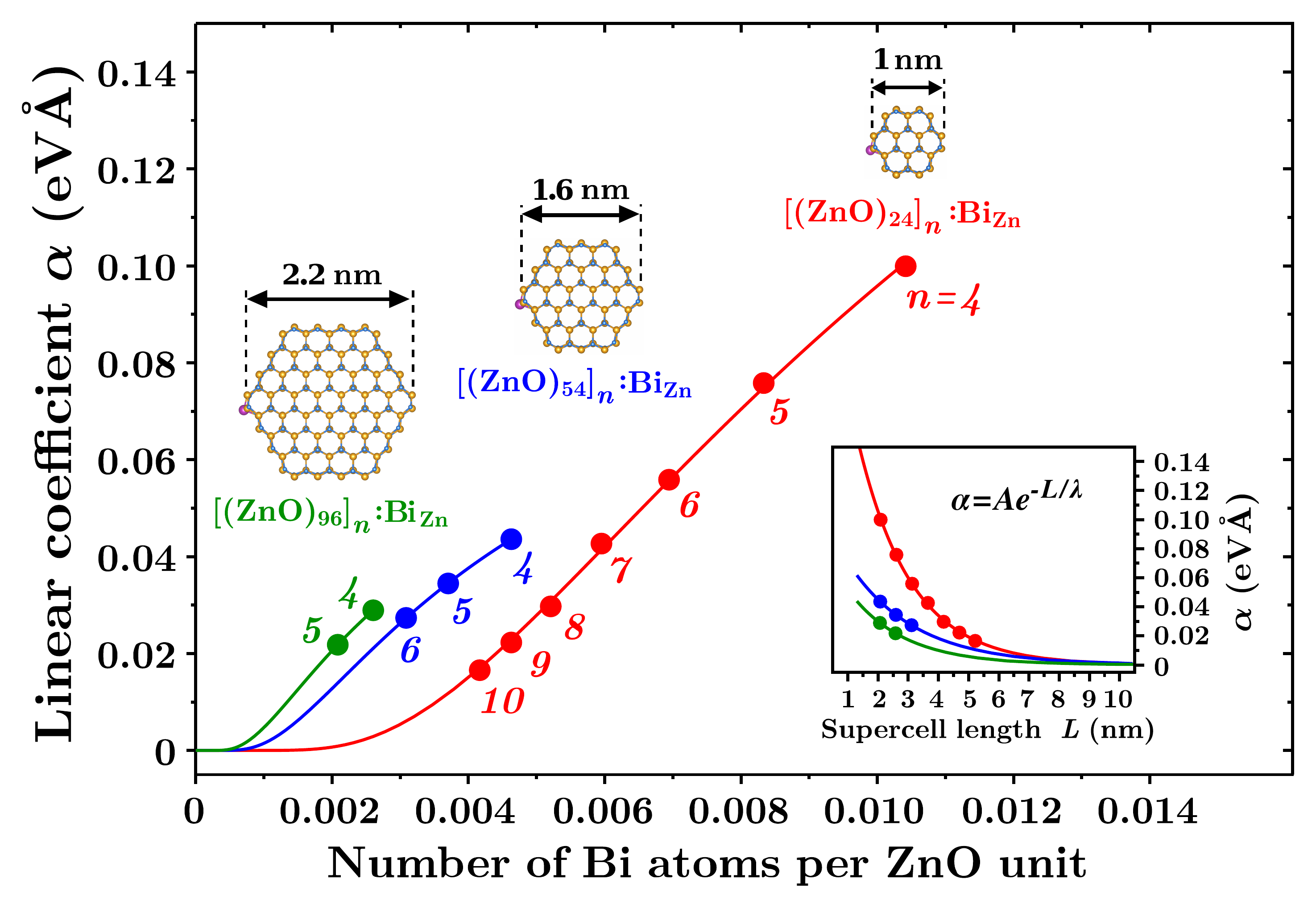}
  \caption{The variation of the linear coefficient $\alpha$
           introduced in Eq.~(\ref{e:disp})
           with the number of Bi atoms per ZnO unit and
          (the inset) the supercell length $L$ along the wire axis.
          }
  \label{f:alfa}
\end{figure}

\begin{figure*}
\centering
  \includegraphics[width=0.964\textwidth]{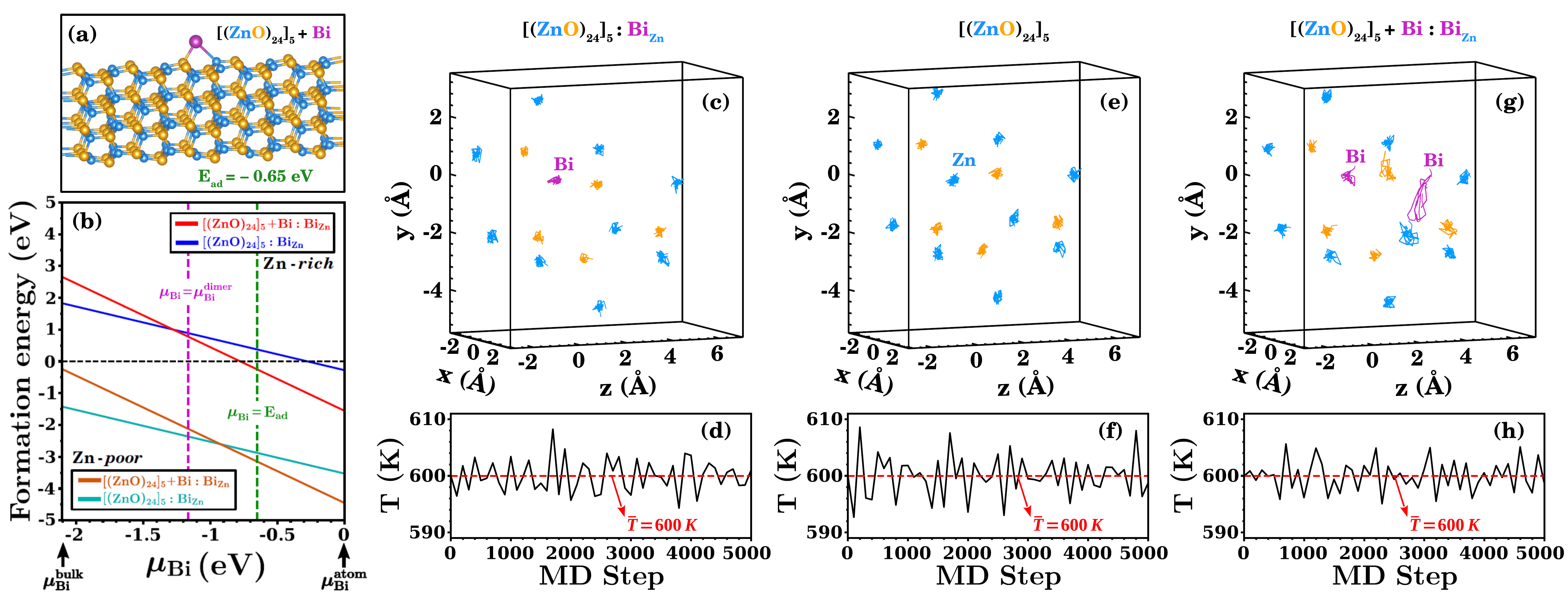}
  \caption{(a) Optimized atomic structure for 
               $\text{[(ZnO)}_{24}\text{]}_5\text{+Bi}$.
           (b) The formation energies of $\text{[(ZnO)}_{24}\text{]}_5\text{:Bi}_{\text{Zn}}$ and
                                         $\text{[(ZnO)}_{24}\text{]}_5\text{+Bi}\text{:Bi}_{\text{Zn}}$
             as a function of the Bi chemical potential $\mu_{\text{\tiny Bi}}$
             under Zn--\textit{rich} and Zn--\textit{poor} conditions;
             the zero of $\mu_{\text{\tiny Bi}}$ is set to $E(\text{Bi})$.
           [(c),(e), and (g)] The trajectories of atoms for
             (c) $\text{[(ZnO)}_{24}\text{]}_5\text{:Bi}_{\text{Zn}}$
             (e) $\text{[(ZnO)}_{24}\text{]}_5$, and
             (g) $\text{[(ZnO)}_{24}\text{]}_5\text{+Bi}\text{:Bi}_{\text{Zn}}$,
             attained from the position of atoms averaged over every 100 MD steps,
             in the course of \textit{ab initio} MD simulations;
             only \textit{five} O and \textit{ten} Zn atoms of the host nanowire are shown,
             which correspond to the first- and second-neighbors of the \textit{two} Bi atoms in (g). 
           [(d),(f), and (h)] The respective graphs showing the temperature $T$ (in black) versus MD steps,
             where the thermostat temperature $\bar{T}$ is marked in red.
          }
  \label{f:stabi}
\end{figure*}

In order to see \textit{if} doping a ZnO nanowire with elements other than Bi
  causes a spin splitting along the momentum axis as in Fig.~\ref{f:cb}(c),
  we repeated geometry optimizations and electronic structure calculations
  by replacing Bi with Tl, Au, or Sb.
Figures~\ref{f:s2}(a)-\ref{f:s2}(d) display
  the electronic energy bands calculated for
  $\text{[(ZnO)}_{24}\text{]}_5\text{:Bi}_{\text{Zn}}$,
  $\text{[(ZnO)}_{24}\text{]}_5\text{:Sb}_{\text{Zn}}$,
  $\text{[(ZnO)}_{24}\text{]}_5\text{:Tl}_{\text{Zn}}$ and
  $\text{[(ZnO)}_{24}\text{]}_5\text{:Au}_{\text{Zn}}$, respectively,
  which are colored to reflect
  the percent contribution from the dopant (Bi, Sb, Tl, or Au) to the electronic states.
The coloring is accomplished by computing
  the contributions from the dopant and host (Zn and O) atoms that are obtained
  by projecting the state wave functions onto spherical harmonics within a sphere around each atom.
We see that the SO splitting in $\text{[(ZnO)}_{24}\text{]}_5\text{:Sb}_{\text{Zn}}$
  is virtually negligible,
  which is as anticipated because the SO effects are much less pronounced for a lighter (Sb) atom
  compared to a heavier (Bi) atom.
We find in fact that
  the value of $\alpha$ computed for $\text{[(ZnO)}_{24}\text{]}_5:\text{Sb}_{\text{Zn}}$
  is \textit{an order of magnitude}
  smaller than the respective value for $\text{[(ZnO)}_{24}\text{]}_5\text{:Bi}_{\text{Zn}}$.
On the other hand,
  doping with heavy elements Tl or Au
  results in a \textit{nonzero} magnetic moment
  owing to formation of \textit{gap states} with {\it flat} dispersion,
  cf. Figs.~\ref{f:s2}(c) and \ref{f:s2}(d).
Thus, the SO splitting of CBM in
  $\text{[(ZnO)}_{24}\text{]}_5:\text{Tl}_{\text{Zn}}$ and
  $\text{[(ZnO)}_{24}\text{]}_5:\text{Au}_{\text{Zn}}$
  occurs along the energy (as opposed to momentum) axis.

Figures~\ref{f:alfa} and S2 (see Ref.~\onlinecite{supmat} for Supplemental Material) show the variation of the linear coefficient $\alpha$ introduced in Eq.~(\ref{e:disp})
  with the Bi concentration for a set of host nanowires $\text{[}\text{(ZnO)}_{N}\text{]}_n$ with increasing thicknesses
  corresponding to $N=$ 24, 54, and 96.
For each value of $N$,
  a parameterization of $\alpha$ as a function of the supercell length $L$ along the wire axis was performed,
  introducing two parameters $A$ and $\lambda$,
  which is shown in the inset of Fig.~\ref{f:alfa}.
The latter is used to plot $\alpha$ versus the \textit{number of Bi atoms per ZnO unit} [Fig.~\ref{f:alfa}] and 
                                           the Bi concentration in units of cm$^{-1}$ [Fig.~S2 (see Ref.~\onlinecite{supmat} for Supplemental Material)].
We see that $\alpha$ tends to reduce as the Bi concentration diminishes,
  which is in line with the fact that the linear-in-$k$ splitting does \textit{not} occur in the \textit{absence} of the dopant.
Furthermore, the thicker the host nanowire the smaller the $\alpha$ value in Fig.~\ref{f:alfa}.
It is thus clear that the value of $\alpha$, which manifests the degree of linear-in-$k$ splitting,
  could be tuned by adjusting the dopant concentration.
The latter can be achieved by \textit{either} decreasing the amount of the dopants (for a given ZnO nanowire)
                              \textit{or} increasing the number of Zn--O pairs, \textit{i.e.}, using a thicker nanowire.
Consequently, doping-induced spin splitting in Bi-doped ZnO nanowires
  could be tuned and controlled by
  not only adjusting the amount of the dopants but also choosing a host nanowire of an adequate thickness.

Bismuth has a low solubility in zinc oxide,
  which is known to give rise to the Bi \textit{segregation} observed in ZnO varistors\cite{smith89,kobayashi98}.
In accordance with this,
  our theoretical characterization\cite{kilic16} of Bi-doped ZnO nanowires in a site-specific manner
  (as regards the location and charge-state of the dopant)
  showed that
  the dopant (Bi) atoms are predominantly substituted into the Zn sites on the nanowire surface.
This implies that
  the doping configuration displayed in Fig.~\ref{f:cb}(a) would abound in the defect structure of
  Bi-doped ZnO nanowires.
Here we employ defect calculations
  to see \textit{if} the configurations displayed in Figs.~\ref{f:cb}(a) and \ref{f:cb}(b)
  could be \textit{realized}
  by putting the host nanowire into contact with a reservoir of bismuth.
Thus, the formation energy of the Bi dopant in these configurations
  is plotted as a function of the Bi chemical potential $\mu_{\text{\tiny Bi}}$ in Fig.~\ref{f:stabi}(b)
  under Zn--\textit{rich} as well as Zn--\textit{poor} conditions.
The formation of an adsorbed Bi atom,
  denoted as $\text{[(ZnO)}_{24}\text{]}_5\text{+Bi}$,
  is also considered,
  for which the lowest-energy configuration is shown in Fig.~\ref{f:stabi}(a).
The values of $\mu_{\text{\tiny Bi}}$
  corresponding to an equilibrium of
  the foregoing doping configurations
  with a reservoir consisting of
    (i) the Bi monomer gas,
   (ii) the Bi$_2$ dimer gas\cite{zajaczkowski16},
  (iii) the adsorbed Bi atoms, or
   (iv) the Bi solid
  are marked in Fig.~\ref{f:stabi}(b).
It is seen that
  the formation energies of
  $\text{[(ZnO)}_{24}\text{]}_5\text{:Bi}_{\text{Zn}}$ and
  $\text{[(ZnO)}_{24}\text{]}_5\text{+Bi}\text{:Bi}_{\text{Zn}}$
  are both \textit{negative}
  under Zn--poor (\textit{i.e.}, O--rich) conditions,
  \textit{regardless} the bismuth reservoir.
Hence these doping configurations
  can clearly be \textit{realized}
  by putting the host nanowire into contact with an adequate reservoir of Bi
  under controlled thermodynamic conditions.

The \textit{stability} of the foregoing doping configurations, cf. Figs.~\ref{f:cb}(a) and \ref{f:cb}(b),
  is examined, in a \textit{relative} manner,
  by performing \textit{ab initio} MD simulations
  at temperature 600~K.
The trajectories of atoms for
             $\text{[(ZnO)}_{24}\text{]}_5\text{:Bi}_{\text{Zn}}$,
             $\text{[(ZnO)}_{24}\text{]}_5$, and
             $\text{[(ZnO)}_{24}\text{]}_5\text{+Bi}\text{:Bi}_{\text{Zn}}$
             attained from the position of atoms (averaged over every 100 MD steps)
  are displayed in Figs.~\ref{f:stabi}(c), \ref{f:stabi}(e), and \ref{f:stabi}(g), respectively.
The respective graphs of the temperature $T$ (also averaged over every 100 MD steps)
  versus MD steps
  are given in Figs.~\ref{f:stabi}(d), \ref{f:stabi}(f), and \ref{f:stabi}(h),
  which show that
  the temperature fluctuations are restricted to an interval of 20~K
  about the thermostat temperature $\bar{T}=600$~K.
The duration of our MD simulations is 5~ps (excluding the equilibration stage of 1 ps),
  which is likely shorter than needed
  for a complete prediction of the dynamical behavior of the studied systems.
Nevertheless, 
  it is clear from Figs.~\ref{f:stabi}(c), \ref{f:stabi}(e), and \ref{f:stabi}(g) that
  not only the host (Zn, O) atoms but also the Bi dopant and adatom
  show a tendency to restore their equilibrium positions
  at a temperature (600 K) considerably higher than room temperature.
Whereas the Bi adatom [Fig.~\ref{f:stabi}(g)] has the \textit{greatest} displacement,
  the Bi dopant [Fig.~\ref{f:stabi}(c and g)] has the \textit{least} displacement
  in comparison to the displacements of the host (Zn, O) atoms.
This indicates that the Bi-doped ZnO nanowire is in fact 
  \textit{as stable as} the undoped ZnO nanowire,
  which is also supported by 
  an overall comparison of the trajectory plots in Figs.~\ref{f:stabi}(c) and \ref{f:stabi}(e).
Furthermore, although our MD simulations were not intended to study
  the diffusion characteristics of Bi species on the surface of the ZnO nanowire,
  it is clear from Fig.~\ref{f:stabi}(g) that
  the presence of the Bi adatom does \textit{not} harm the stability of the \textit{substitutional} Bi dopant.

\section{\label{netice}Conclusion}

The present investigation demonstrates that
  doping a semiconducting nanowire with a heavy element could be an effective means to lift Kramers degeneracy
  of the conduction-band states.
In particular, doping ZnO nanowires with Bi
  is identified as a means to design quasi-one-dimensional materials for spintronic applications,
  thanks to the occurrence of the linear-in-$k$ spin-orbit splitting explored in this study.
It is noteworthy that
  the degree of the linear-in-$k$ spin splitting,
  which would facilitate
  the control of spin-polarization of the conduction electrons in Bi-doped ZnO nanowires
  by applying external \textit{electric} fields,
  could be tuned by adjusting the dopant concentration.
Our findings suggest that Bi-doped ZnO nanowires,
  which are found to be stable under reasonable thermodynamic conditions,
  could be used in spintronic applications.

\begin{acknowledgments}
The authors acknowledge financial support from TUBITAK through Grant 114F155.
The numerical calculations reported here were carried out 
  at the High Performance and Grid Computing Center (TRUBA Resources) of TUBITAK ULAKBIM.
\end{acknowledgments}

\appendix*

\section{Lowering of the symmetry in the presence of the dopant}

\begin{figure}
\centering
  \includegraphics[width=0.482\textwidth]{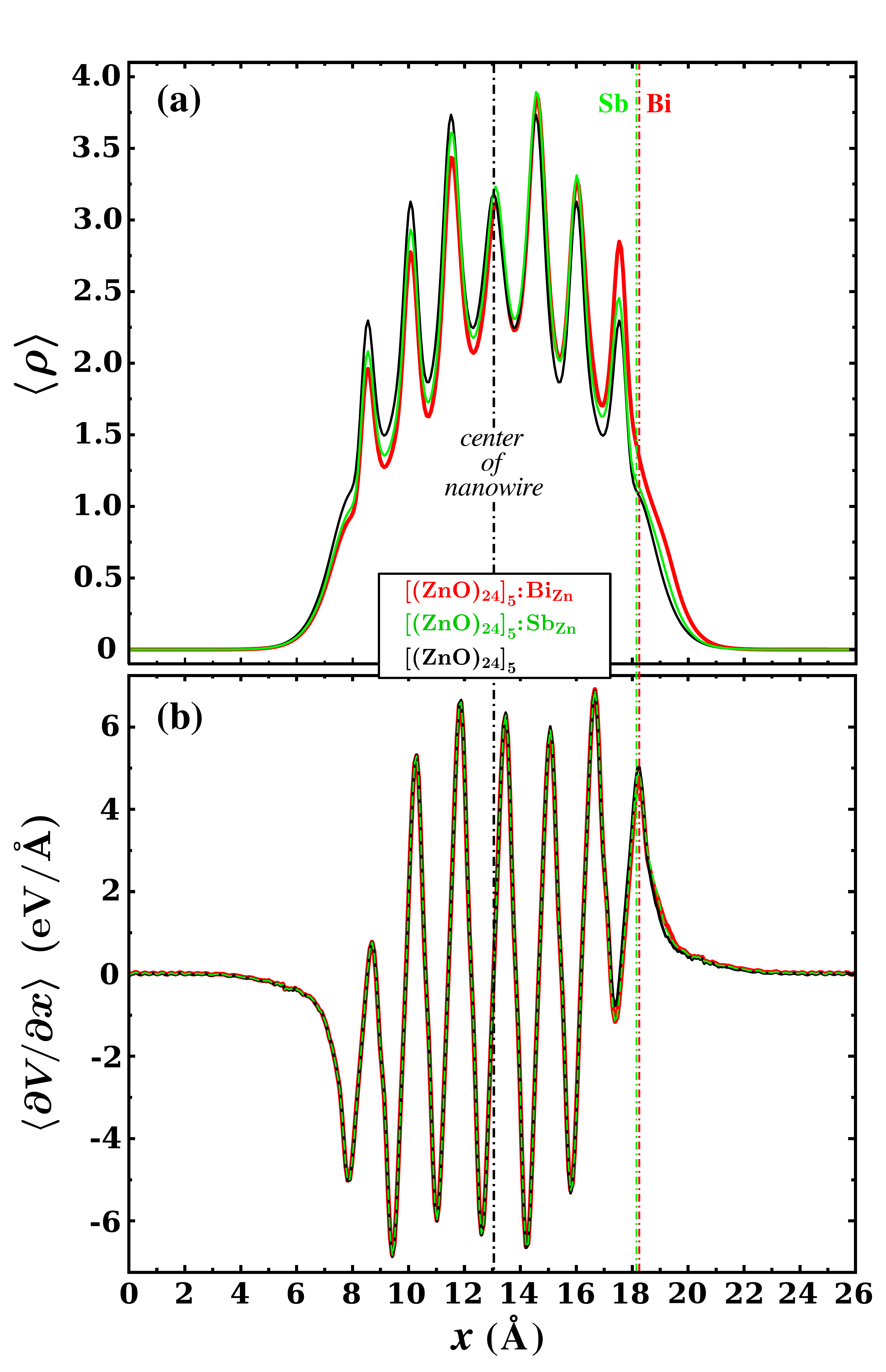}
  \caption{The curves in red, green and black represent
           the plots of
           the $yz$ planar average of
           (a) the state charge density $\langle \rho \rangle$ and
           (b) the potential gradient $\langle \partial V/\partial x \rangle$
           versus $x$,
           cf. Fig.~\ref{f:plane},
           for $\text{[(ZnO)}_{24}\text{]}_5\text{:Bi}_{\text{Zn}}$,
               $\text{[(ZnO)}_{24}\text{]}_5\text{:Sb}_{\text{Zn}}$ and
               $\text{[(ZnO)}_{24}\text{]}_5$, respectively.
           The vertical dash-dot lines mark the center of nanowire (black)
             as well as the positions of Bi (red) and Sb (green) atoms on the $x$ axis.
          }
  \label{f:b}
\end{figure}

From the perspective of symmetry,
  the linear-in-$k$ SO splitting in Eq.~(\ref{e:disp})
  \textit{cannot} be attributed to the host (ZnO) nanowire.
Since the irreducible representations of C$_{6\text{v}}$ compatible with spin are doubly-degenerate,
  \textit{no} spin splitting occurs along the $\Gamma$-A direction of the wurtzite Brillouin zone\cite{voon96}.
This is also valid for the ZnO nanowires employed here, whose point group is C$_{6\text{v}}$, in the \text{absence} of the Bi dopant,
  cf. Fig. S1(a) (see Ref.~\onlinecite{supmat} for Supplemental Material).
It is thus clear that
  the SO splitting explored here
  occurs due to lowering of the symmetry in the presence of the dopant.
Here we present an analysis of the latter 
  in terms of doping-induced changes in 
  the potential gradient and the CBM wave function 
  since the expectation value of the SO interaction operator
  involves \textit{both}, cf. Eq.~(\ref{e:hso}).
This analysis is conducted
  by plotting the $yz$ planar average of
  the state charge density $\rho =\left | \psi_\mathbf k \right |^2$ in Fig.~\ref{f:b}(a) and 
  the potential gradient $\partial V/\partial x$ in Fig.~\ref{f:b}(b)
  with respect to $x$,
  where $\mathbf k$ is set as the wave vector at which the CBM occurs.
It is instructive to compare the Bi-doped nanowire to not only the undoped nanowire
                                                      but also the Sb-doped nanowire
  because $\text{[(ZnO)}_N\text{]}_n\text{:Bi}_{\text{Zn}}$ and 
          $\text{[(ZnO)}_N\text{]}_n\text{:Sb}_{\text{Zn}}$ possess the same symmetry.
As respectively seen in Figs.~\ref{f:b}(a) and \ref{f:b}(b), both
  $\langle \rho \rangle$           and 
  $\langle \partial V/\partial x \rangle$ 
  have asymmetric profiles in the presence of the dopant (Bi or Sb),
  which could be explained by noting that
  the reflection symmetry with the mirror plane at the center of the nanowire
  is lost when the dopant is incorporated.
It is also seen that 
  the doping-induced change in $\langle \partial V/\partial x \rangle$ is not noteworthy
  whereas the change in $\langle \rho \rangle$ shows a significantly asymmetric profile,
  as will be quantified in the next paragraph.
Comparing the curves for
  $\text{[(ZnO)}_N\text{]}_n\text{:Bi}_{\text{Zn}}$ and 
  $\text{[(ZnO)}_N\text{]}_n\text{:Sb}_{\text{Zn}}$ in Fig.~\ref{f:b}(b) to each other,
  it is seen that the change in the potential gradient is quite indifferent to the type of dopant,
  which is seemingly not the case with the state charge density
  since the red curve, in comparison to the green curve, is in a larger deviation from the black (symmetric) curve 
  in Fig.~\ref{f:b}(a).
These comparisons show that the essential difference between doping with Bi and Sb
  pertains mostly to the change in the CBM wave function,
  rather than the potential gradient.
This means that reducing the symmetry of $\vec{\nabla} V$ 
  would \textit{not necessarily} produce a SO splitting as in Eq.~(\ref{e:disp});
  it \textit{must} be accompanied by the adequate modification of the state wave function to a lower symmetry,
  cf. Fig.~\ref{f:b}(a),
  as in the case of $\text{[(ZnO)}_N\text{]}_n\text{:Bi}_{\text{Zn}}$.

The lowering of the symmetry in the presence of the dopant (Bi or Sb)
  can be quantified by introducing the following quantities:
$I      = \int dx \langle \rho        \rangle \langle \frac{\partial V       }{\partial x} \rangle $, 
$I_0    = \int dx \langle \rho_0      \rangle \langle \frac{\partial V_0     }{\partial x} \rangle $, 
$I_V    = \int dx \langle \rho        \rangle \langle \frac{\partial \Delta V}{\partial x} \rangle $, and 
$I_\rho = \int dx \langle \Delta \rho \rangle \langle \frac{\partial V       }{\partial x} \rangle $,
where
  $\Delta V(\mathbf r)=V(\mathbf r) - V_0(\mathbf r)$ and
  $\Delta \rho(\mathbf r)=\rho(\mathbf r) - \rho_0(\mathbf r)$, and
  $\rho(\mathbf r)$   and $V(\mathbf r)$ 
 [$\rho_0(\mathbf r)$ and $V_0(\mathbf r)$] denote the state charge density and potential of the doped [undoped] nanowire.
Note that   (i) $I=I_0+I_V+I_\rho$;
  the computed value of $I = 0.08$~eV for $\text{[(ZnO)}_{24}\text{]}_5\text{:Sb}_{\text{Zn}}$
  is \textit{half} of that for $\text{[(ZnO)}_{24}\text{]}_5\text{:Bi}_{\text{Zn}}$ for which $I = 0.16$~eV.
           (ii) $I_0$ involves only the host-related quantities, and so $I_0 = 0$ owing to the symmetry of the undoped ZnO nanowire. 
          (iii) $I_V$ and 
                $I_\rho$ are ``measures'' of the degree of doping-induced asymmetry due to the change in the potential and 
                                                                                                      in the state charge density, respectively.
We obtain $I_V=-0.03$~eV for \textit{both} $\text{[(ZnO)}_{24}\text{]}_5\text{:Bi}_{\text{Zn}}$ and
                                           $\text{[(ZnO)}_{24}\text{]}_5\text{:Sb}_{\text{Zn}}$
  whereas $I_\rho=0.19$ and 0.11~eV for $\text{[(ZnO)}_{24}\text{]}_5\text{:Bi}_{\text{Zn}}$ and
                                        $\text{[(ZnO)}_{24}\text{]}_5\text{:Sb}_{\text{Zn}}$,respectively.
Note that $I_V$ take the same value for both dopants, which is significantly smaller than the respective $I_\rho$ values.
This means that the doping-induced asymmetry due to the changes in the state wave function,
   rather than in the potential gradient, is pronounced.
Accordingly, what really differentiates the cases of doping Bi versus Sb is the lowering of the symmetry of the CBM wave function.

\clearpage

\section*{Supplemental Material}

\setcounter{page}{1}
\setcounter{figure}{0}
\setcounter{table}{0}
\renewcommand{\thepage}{S-\arabic{page}}
\renewcommand{\thefigure}{S\arabic{figure}}
\renewcommand{\thetable}{S\arabic{table}}

\begin{itemize}
\item Figure~S1 displays the electronic energy bands calculated for
        $\text{[(ZnO)}_{24}\text{]}_5$,
        $\text{[(ZnO)}_{24}\text{]}_5\text{:Bi}_{\text{Zn}}$,
        $\text{[(ZnO)}_{24}\text{]}_5\text{+Bi}\text{:Bi}_{\text{Zn}}$,
        $\text{[(ZnO)}_{24}\text{]}_5\text{+Bi}$,
        which are colored according to the contribution of atoms.
\item Figure~S2 shows the plot of the linear coefficient $\alpha$
        versus the Bi concentration in units of cm$^{-1}$.
\end{itemize}

\clearpage

\begin{figure*}
  \begin{center}
    \resizebox{0.723\textwidth}{!}{
      \includegraphics{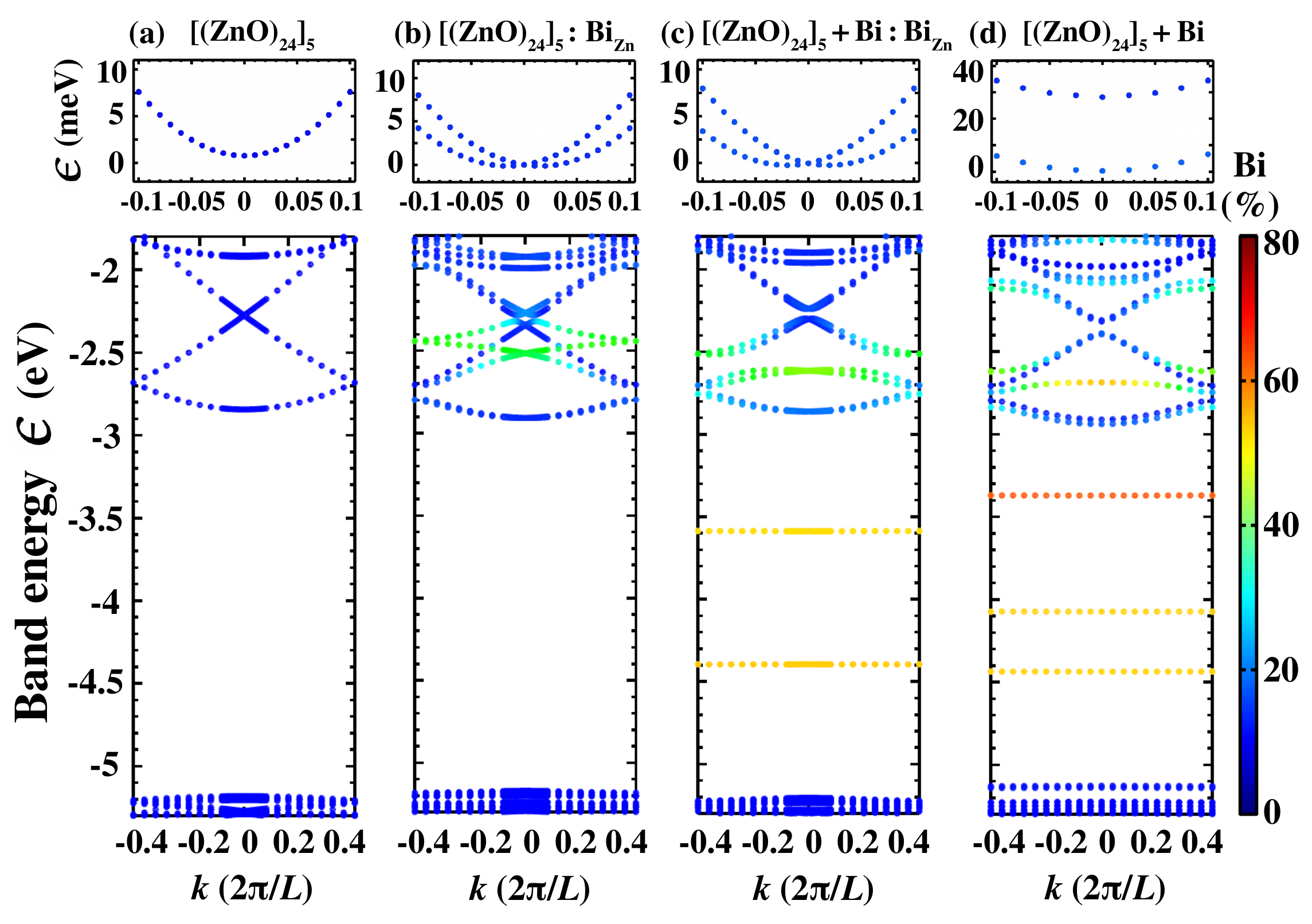}
    }
  \end{center}
  \vspace*{-0.75cm}
  \caption{The electronic energy bands calculated for
           (a) $\text{[(ZnO)}_{24}\text{]}_5$,
           (b) $\text{[(ZnO)}_{24}\text{]}_5\text{:Bi}_{\text{Zn}}$,
           (c) $\text{[(ZnO)}_{24}\text{]}_5\text{+Bi}\text{:Bi}_{\text{Zn}}$, and
           (d) $\text{[(ZnO)}_{24}\text{]}_5\text{+Bi}$.
           The circles are colored to reflect 
           the percent contribution from Bi to the electronic states.
           The latter is computed by using
           the contributions from Bi, Zn, and O atoms that are obtained
           by projecting the state wave functions onto spherical harmonics within a sphere around each atom.
           The upper panels show a close-up view of the \textit{two} lowest conduction bands.
          }
\end{figure*}

\begin{figure}
  \begin{center}
    \resizebox{0.482\textwidth}{!}{
      \includegraphics{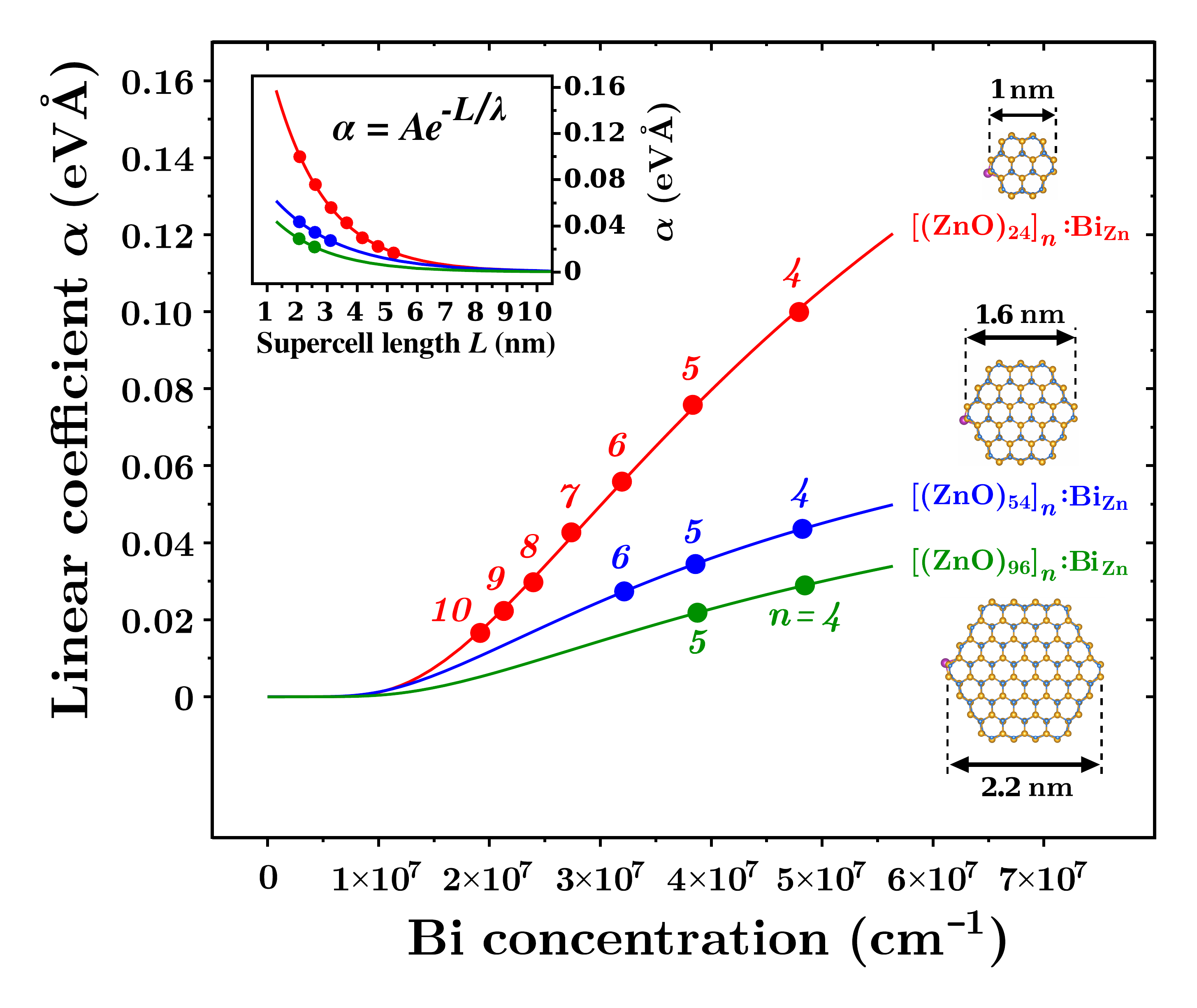}
    }
  \end{center}
  \vspace*{-0.75cm}
  \caption{The variation of the linear coefficient $\alpha$
           with the Bi concentration and
          (the inset) the supercell length $L$ along the wire axis.
          }
\end{figure}

\end{document}